# Searching for Propionamide ($C_2H_5CONH_2$) Toward Sagittarius B2 at Centimeter Wavelengths

Caden Schuessler,[1] Anthony Remijan,[2] Ci Xue,[3] Joshua Carder,[1] Haley Scolati,[1] and Brett A. McGuire[3,2]

[1]*Department of Chemistry, University of Virginia, Charlottesville, VA 22903, USA*
[2]*National Radio Astronomy Observatory, Charlottesville, VA 22903, USA*
[3]*Department of Chemistry, Massachusetts Institute of Technology, Cambridge, MA 02139, USA*

## ABSTRACT

The formation of molecules in the interstellar medium (ISM) remains a complex and unresolved question in astrochemistry. A group of molecules of particular interest involves the linkage between a -carboxyl and -amine group, similar to that of a peptide bond. The detection of molecules containing these peptide-like bonds in the ISM can help elucidate possible formation mechanisms, as well as indicate the level of molecular complexity available within certain regions of the ISM. Two of the simplest molecules containing a peptide-like bond, formamide ($NH_2CHO$) and acetamide ($CH_3CONH_2$), have previously been detected toward the star forming region Sagittarius B2 (Sgr B2). Recently, the interstellar detection of propionamide ($C_2H_5CONH_2$) was reported toward Sgr B2(N) with ALMA observations at millimeter wavelengths. Yet, this detection has been questioned by others from the same set of ALMA observations as no statistically significant line emission was identified from any uncontaminated transitions. Using the PRrbiotic Interstellar MOlecule Survey (PRIMOS) observations, we report an additional search for $C_2H_5CONH_2$ at centimeter wavelengths conducted with the Green Bank Telescope. No spectral signatures of $C_2H_5CONH_2$ were detected. An upper limit for $C_2H_5CONH_2$ at centimeter wavelengths was determined to be $N_T < 1.8 \times 10^{14}$ cm$^{-2}$ and an upper limit to the $C_2H_5CONH_2/CH_3CONH_2$ ratio is found to be <2.34. This work again questions the initial detection of $C_2H_5CONH_2$ and indicates that more complex peptide-like structures may have difficulty forming in the ISM or are below the detection limits of current astronomical facilities. Additional structurally related species are provided to aid in future laboratory and astronomical searches.

*Keywords:* Astrochemistry, ISM: molecules

## 1. INTRODUCTION

Proteins are essential for mediating and maintaining life on Earth. All proteins consist of amino acids connected by peptide bonds formed between respective carboxyl (R-COOH) and amine (a N-atom with a singular pair of electrons) groups. As proteins are a requirement of terrestrial life, they are of particular interest within the cosmic environment. Given their high molecular weight, complexity, and low gas-phase abundance, the detection of proteins in the interstellar medium (ISM) is unlikely at this point in time. Instead of searching for these biomolecules directly, a more tractable approach is to look for structurally similar peptide-like species (e.g. Colzi et al. 2021). These molecules have the same basic structure but with smaller and less complex side chains. Although these peptide-like molecules are of particular interest, only a limited number have been detected.

The detection of such molecules may be maximized by searching in chemically active regions. Sagittarius B2 (Sgr B2) is a high mass star forming region abundant in molecular gas and dust near the Galactic center. The region is complex, having sources with varying densities and temperatures (Candelaria et al. 2021). The simplest peptide-like molecule, formamide ($NH_2CHO$), was initially detected toward Sgr B2 (Rubin et al. 1971) and has since been observed toward numerous sources (e.g. Blake et al. 1986; Bisschop et al. 2007; Adande et al. 2013; Belloche et al. 2019). The detection of $NH_2CHO$ has garnered extensive attention recently as it is a possible chemical precursor to the formation of larger, more complex and important biological molecules (López-Sepulcre et al. 2019). Acetamide ($CH_3CONH_2$), the next simplest peptide-like molecule, has also been detected toward Sgr B2(N) in multiple surveys (e.g. Hollis et al. 2006; Halfen et al. 2011; Remijan et al. 2022) and towards G31.41+0.31 (Colzi et al. 2021). Other chemically related molecules, such as N-methylformamide ($CH_3NHCHO$) and urea (($NH_2)_2CO$) in Sgr B2(N) and G+0.693-0.027 (Jiménez-Serra et al. 2020), as well as $CH_3NHCHO$ in NGC 6334I and G31.41+0.31, have since been detected (Colzi et al. 2021; Belloche et al. 2017; Ligterink et al. 2020). These findings suggest that peptide-like species and other chemically related molecules may be diverse and widespread in space (Belloche et al. 2017; Colzi et al. 2021).

To further this investigation, following on from the detections of $NH_2CHO$ and $CH_3CONH_2$, a larger peptide-like molecule that can be searched for is propionamide ($C_2H_5CONH_2$). However, its identification in the ISM has proved challenging - while Li et al. (2021) initially claimed a detection toward two regions in Sgr B2(N), this detection has recently been challenged by Kolesniková et al. (2022) with a similar set of millimeter observations. In this work, we report an independent search for $C_2H_5CONH_2$ toward Sgr B2 at centimeter wavelengths. The observations of several large molecules, including $CH_3CONH_2$, have been successful at centimeter wavelengths especially with transitions exhibiting maser action. Unfortunately, no transitions of $C_2H_5CONH_2$ were detected. The observing parameters describing the search for $C_2H_5CONH_2$ toward Sgr B2(N) are presented in Section 2. The results of the search for $C_2H_5CONH_2$, and the previously detected peptide-like molecules $NH_2CHO$ and $CH_3CONH_2$, are shown in Section 3. A discussion of the formation routes and the possibility of searching for additional structurally similar species is given in Section 4. Finally, our conclusions are highlighted in Section 5.

## 2. OBSERVATIONS

The PRebiotic Interstellar MOlecule Survey (PRIMOS)[1] was a key science program that started in 2008 January and concluded in 2011 July conducted with the Robert C. Byrd Green Bank Telescope (GBT) that is currently managed by the Green Bank Observatory. The primary goal of PRIMOS was to perform the deepest spectral line survey to date toward Sgr B2(N) at centimeter wavelengths.

---

[1] Access to the entire PRIMOS data set, specifics on the observing strategy, and overall frequency coverage information is available at http://archive.nrao.edu by searching for GBT Program ID: AGBT07A_051

These data afforded the detection of many molecular species, and specifically those molecules important in prebiotic formation pathways and novel formation or excitation mechanisms. For example, the publicly accessible PRIMOS data were used to detect trans-methyl formate ($CH_3OCHO$) (Neill et al. 2012), carbodiimide (HNCNH) (McGuire et al. 2012), and ethanimine ($CH_3CHNH$) (Loomis et al. 2013).

The PRIMOS project covers nearly all observable frequencies available to ground-based instrumentation from ∼300 MHz to 48 GHz at high sensitivity (∼3–9 mK rms) and spectral resolution (24.4 kHz) toward the source Sgr B2(N). The observations were centered on the Sgr B2(N) "Large Molecule Heimat" (LMH) at $(\alpha, \delta)_{J2000}=(17^h47^m19\overset{s}{.}8, -28°22'17\overset{''}{.}0)$. This source was chosen as it is the most prominent location for studying the abundance of complex interstellar species, including the first detections of interstellar acids (Winnewisser & Churchwell 1975; Liu et al. 2001; Mehringer et al. 1997) and amide species (Rubin et al. 1971; Hollis et al. 2006). About half of all identified interstellar molecules were first reported toward this source (McGuire 2022), including possible precursor molecules for $C_2H_5CONH_2$: $NH_2CHO$ and $CH_3CONH_2$.

## 3. RESULTS

### 3.1. Absorption Lines of $NH_2CHO$

Data taken as part of PRIMOS have been used in the detection of several new molecular species in the ISM as well in uncovering the physical and chemical environments of the extended envelope surrounding the Sgr B2(N-LMH) hot molecular core. To start the investigation into the possible detection of $C_2H_5CONH_2$, a more thorough analysis was conducted on the structurally similar species $NH_2CHO$ and $CH_3CONH_2$. Hollis et al. (2006) reported the first detection of interstellar $CH_3CONH_2$ and in that work, presented two transitions of $NH_2CHO$. The two $NH_2CHO$ transitions presented by Hollis et al. (2006) showed strong absorption components at 21207 MHz at $v_{LSR}$ velocities of +64 and +82 km s$^{-1}$; however, the transition at 9237 MHz was detected in emission with both resolved hyperfine structure and with $v_{LSR}$ components at +64 and +82 km/s. Neglecting the possibility of non-thermal excitation to account for the emission features, Hollis et al. (2006) reported a column density range between 1-90×$10^{14}$ cm$^{-2}$ at an excitation temperature of ∼5 K. It is now well known that many of the transitions detected from these low frequency observations on the GBT can only be explained by maser activity (e.g. McGuire et al. 2012; Faure et al. 2014) and that the reported column densities based on thermal excitation can only be considered approximate. In addition, highly abundant species such as $NH_2CHO$ are not necessarily confined to a singular physical environment and are found in regions from compact hot cores to more extended cooler envelopes (Corby et al. 2015; Thiel et al. 2017; Belloche et al. 2017). These low frequency transitions detected with the GBT can also be optically thick, and in general do not strictly follow thermal excitation - thus making the modeling of their spectrum difficult.

While several transitions of $NH_2CHO$ were detected in emission in the PRIMOS data, such as the 9237 MHz feature reported by (Hollis et al. 2006), these emission lines are not shown nor used to constrain the physical environment in this study. It is beyond the scope of this work to fully account for the non-thermal excitation needed to simulate the maser activity from these transitions. Figure 1 shows four transitions of $NH_2CHO$ that were detected in absorption in the PRIMOS data. Figure 1a shows the detection of $NH_2CHO$ at 21207 MHz as originally reported by Hollis et al. (2006). Table 2 summarizes the spectroscopic information for the $NH_2CHO$ lines investigated in this work.

The MOLSIM package (Lee et al. 2021) was used to simulate the $NH_2CHO$ absorption transitions detected in the PRIMOS data under the single-excitation-temperature assumption. The formalism for how to fit for physical parameters of a source (e.g. excitation temperature and column density) from absorption line features, and specifically from GBT observations is further described in (Hollis et al. 2004a) and references therein. Five parameters including the velocity structure, line width, excitation temperature, source size, and column density are needed to generate the synthetic spectra. The three velocity components of Sgr B2 were set to the previously observed values (Corby et al. 2015). Two components have a similar source velocity of +63 km s$^{-1}$ and correspond to the H II region K6 and the LMH hot core, respectively. The third component, the H II region K5, has a source velocity of +82 km s$^{-1}$. The positions for these sources can be found in Table 3 of Corby et al. (2015). The best fit line width was set at 10 km s$^{-1}$. The excitation temperature ($T_{ex}$) was set to 5.8 K based on previous best-fit temperatures from other structurally similar and chemically related species, $CH_3CONH_2$ (Remijan et al. 2022).

We had assumed that the molecular gas which absorbs against the background continuum emission has a similar distribution to the continuum emission. However, based on interferometric observations at centimeter wavelengths, the K5 component has an equivalent source size of 11″, the LMH region has an equivalent source size of 3″, and the K6 component has an equivalent source size of 14″. The GBT beam size, and therefore the sources and sizes within the beam, are frequency dependent. While both the K5 and LMH regions are within the beam for the four $NH_2CHO$ transitions, at 40 GHz, the smaller beam results in most of the K6 components lying outside of the beam area. Therefore, we fixed the source size for the K5 and LMH regions while varying the source size of K6 based on the amount of the K6 region that was within the beam at a given observing frequency in order to properly simulate the absorption line profiles of $NH_2CHO$.

As can be seen in Figure 1, the model fit (red trace) was optimized via visual examination to match the PRIMOS data (black trace) over these three velocity components. The optimization was done by adjusting the column densities for the three velocity components to best match the observed PRIMOS transitions. The best fit column density derived from the four absorption lines detected in PRIMOS yields $N_T = 8.5 \times 10^{14}$ cm$^{-2}$ which is a sum over all three velocity components which is a similar process used to determine the total column density of $CH_3CONH_2$ found in (Remijan et al. 2022). The only changes between the 2 methods is 1) the observed antenna temperatures for the background emission are the same but the source size of the continuum is different in the current analysis and 2) the velocity components of $NH_2CHO$ were slightly different than $CH_3CONH_2$. The measured column density is within the range found by Hollis et al. (2006) based on the different temperatures used in that work. Until sufficient modeling is performed to account for the emission detected by maser activity of the other detected $NH_2CHO$ lines, this will remain the best column density determination at $T_{ex}$ = 5.8 K to characterize the spectrum of $NH_2CHO$ at frequencies observed with the GBT.

Using the recent column density determination of $CH_3CONH_2$ by Remijan et al. (2022), the $CH_3CONH_2$/$NH_2CHO$ ratio determined from the PRIMOS survey is 0.09. From Belloche et al. (2019), the determined $CH_3CONH_2$/$NH_2CHO$ ratio of 0.05 was determined for the source Sgr B2(N2) and 0.14 for the Sgr B2(N1S) source. It should be noted here that while the GBT beam does include both the N1 and N1S cores detected by ALMA, given the affects of beam dilution and the large energy differences in the transitions searched for between the two facilities, the comparison of the column density ratios between the detected molecules is comparing the ratios found in two different physical environments - and yet, they are quite similar.

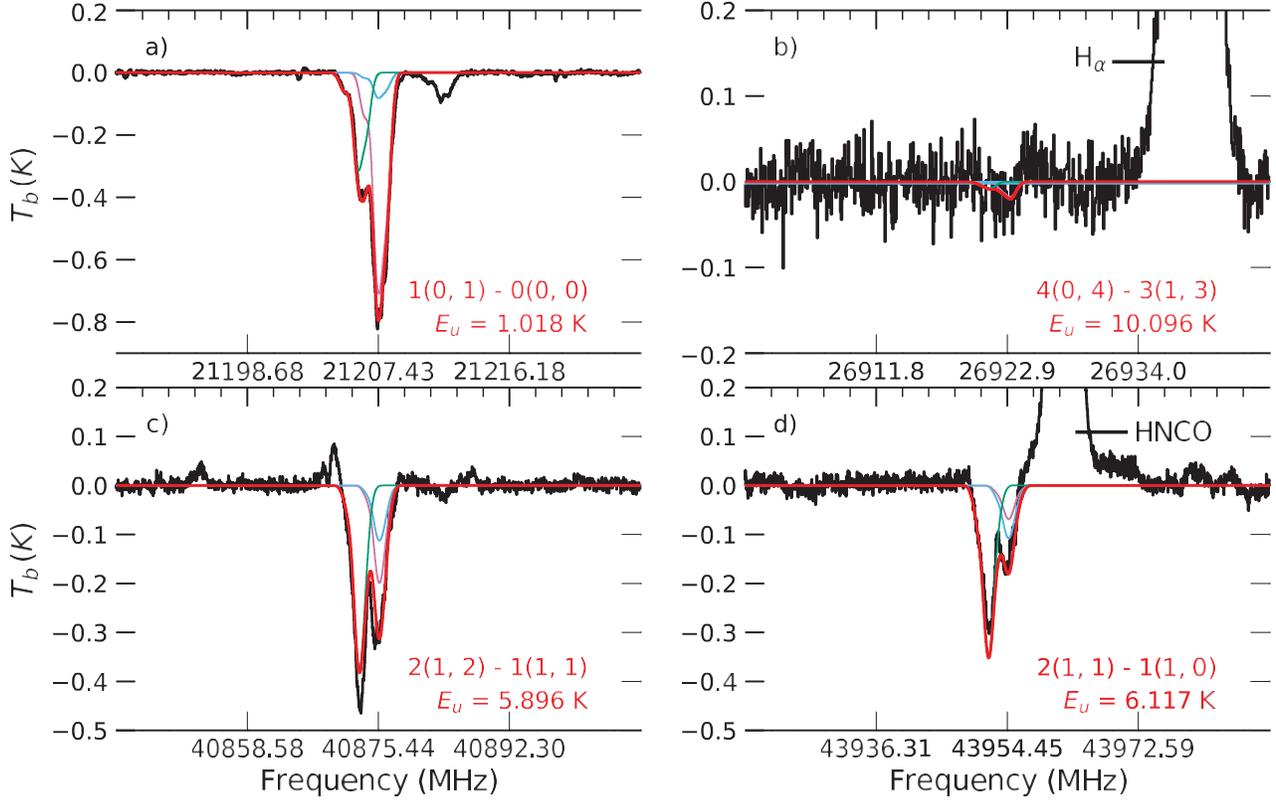

**Figure 1.** Absorption features of NH$_2$CHO detected in the PRIMOS data (black trace). The best-fit model to the PRIMOS observations toward Sgr B2(N), including all velocity components, are shown with red traces with a best-fit $T_{ex}$ of 5.8 K and $N_T$ of 8.5 10$^{14}$ cm$^{-2}$. Simulated spectra of the three velocity components are show in: blue (the LMH region with a velocity of +63 km s$^{-1}$ and a source size of 3″), purple (the K6 region with a velocity of +63 km s$^{-1}$ and a frequency-dependant source size), and cyan (the K5 region with a velocity of +82 km s$^{-1}$ and a source size of 11″), as described in Section 3.1. Transition quantum numbers and upper state energy levels (in K) are at the bottom of each spectrum.

### 3.2. Searching for C$_2$H$_5$CONH$_2$ with the PRIMOS observations

Following this methodology, we searched for transitions of C$_2$H$_5$CONH$_2$ between 8 and 43 GHz in the PRIMOS data, including any possible maser transitions. C$_2$H$_5$CONH$_2$ is one the largest alkyl amides that has been searched for and has been claimed by Li et al. (2021) but seriously questioned by Kolesniková et al. (2022). C$_2$H$_5$CONH$_2$ has dipole moments of $\mu_a$ = 0.6359(60) D and $\mu_b$ = 3.496(30) D, as reported by Marstokk et al. (1996). As such, $b$-type transitions should be the strongest and, as typical for large asymmetric top molecules, $R$-branch transitions should also be the strongest observed transitions (except in those cases where maser activity greatly increases the line intensity).

Since the initial detection of CH$_3$CONH$_2$ (Hollis et al. 2006), interstellar amides (as well as their structural isomers) have been later confirmed by their emission and absorption profiles at both centimeter and millimeter wavelengths (Halfen et al. 2011; Belloche et al. 2019; Remijan et al. 2022;

**Table 1.** Spectroscopic Properties of Simulated Formamide Transitions

| Transition | | Frequency | $E_u$ | $\log_{10}\frac{A_{ul}}{s^{-1}}$ | $S_{ij}\mu^2$ | $T_{bg}$ |
|---|---|---|---|---|---|---|
| $J'_{K'_a,K'_c} - J''_{K''_a,K''_c}$ | $F' - F''$ | (MHz) | (K) | | (D$^2$) | (K) |
| $1_{0,1} - 0_{0,0}$ | 0 − 1 | 21206.4501(30) | 1.018 | -6.315 | 4.358 | 54.1 |
| | 2 − 1 | 21207.3337(5) | | -6.315 | 21.791 | |
| | 1 − 1 | 21207.9218(5) | | -6.315 | 13.076 | |
| $4_{0,4} - 3_{1,3}$ | 3 − 3 | 26921.4054(10) | 10.096 | -8.617 | 0.074 | 35.8 |
| | 4 − 3 | 26922.3900(300) | | -7.550 | 1.116 | |
| | 5 − 4 | 26923.0700(300) | | -7.522 | 1.455 | |
| | 3 − 2 | 26923.3700(300) | | -7.559 | 0.850 | |
| | 4 − 4 | 26923.8312(7) | | -8.726 | 0.074 | |
| $2_{1,2} - 1_{1,1}$ | 1 − 1 | 40874.0048(5) | 5.896 | -5.886 | 4.903 | 17.8 |
| | 1 − 2 | 40874.5663(11) | | -7.062 | 0.327 | |
| | 3 − 2 | 40875.2617(3) | | -5.506 | 27.458 | |
| | 1 − 0 | 40875.4267(7) | | -5.761 | 6.538 | |
| | 2 − 1 | 40875.9331(5) | | -5.631 | 14.712 | |
| | 2 − 2 | 40876.4946(6) | | -6.108 | 4.903 | |
| $2_{1,1} - 1_{1,0}$ | 1 − 0 | 43953.0051(7) | 6.117 | -5.667 | 6.537 | 15.8 |
| | 2 − 2 | 43953.7870(6) | | -6.014 | 4.903 | |
| | 3 − 2 | 43954.3965(3) | | -5.411 | 27.458 | |
| | 1 − 2 | 43954.7321(11) | | -6.968 | 0.327 | |
| | 2 − 1 | 43954.9459(5) | | -5.536 | 14.708 | |
| | 1 − 1 | 43955.8910(5) | | -5.792 | 4.903 | |

NOTE— The spectroscopic data of the NH$_2$CHO transitions corresponding to the observed PRIMOS absorption features are taken from the CDMS (Müller et al. 2005) and the SPLATALOGUE spectroscopy databases [a]. The intrinsic background temperatures were calculated based on the frequency-dependent observed GBT antenna temperatures and source sizes for the Sgr B2 north component as reported in Hollis et al. (2007).

[a] https://www.splatalogue.online

Ligterink et al. 2020; Colzi et al. 2021). The laboratory and spectroscopic information of C$_2$H$_5$CONH$_2$ was reported in Li et al. (2021). From these data, we searched for the strongest transitions of C$_2$H$_5$CONH$_2$ that would be present in the PRIMOS observations - namely the $b$-type and $R$-branch transitions, and for completeness, two $Q$-branch transitions in an attempt to identify any transition that may be excited by maser activity. However, as Figure 2 illustrates, no transitions of C$_2$H$_5$CONH$_2$ were detected beyond the 3$\sigma$ noise level. A weak absorption line is detected in Figure 2c very close to the rest frequency of the 4$_{0,4}$ - 3$_{1,3}$ transition at 21437 MHz. While this transition cannot be assigned to any spectral feature of a known astronomical molecule, it is very unlikely it can be assigned to the 4$_{0,4}$ - 3$_{1,3}$ transition of C$_2$H$_5$CONH$_2$ given the difference in the central frequency of the observed absorption profile in the PRIMOS spectrum and the calculated rest frequency of the C$_2$H$_5$CONH$_2$ transition. It is also very unlikely to be a feature of C$_2$H$_5$CONH$_2$ given that additional transitions at larger (and smaller) $J$ states are not identified. As such, this feature violates several of the "Snyder

**Table 2.** Spectroscopic Properties of Simulated Propionamide Transitions

| Transition | | | Frequency | $E_u$ | $\log_{10}\frac{A_{ul}}{s^{-1}}$ | $S_{ij}\mu^2$ | $T_{bg}$ |
|---|---|---|---|---|---|---|---|
| $J'_{K',K'} - J''_{K'',K''}$ | $F' - F''$ | Sym | (MHz) | (K) | | (D²) | (K) |
| $2_{1,1} - 2_{0,2}$ | 2 − 2 | E | 8072.5029(9) | 1.33 | -8.850 | 1.154 | 304.8 |
| | 2 − 2 | A | 8072.5592(9) | | -8.850 | 1.154 | |
| | 3 − 3 | E | 8073.9129(4) | | -8.741 | 2.073 | |
| | 3 − 3 | A | 8073.9692(4) | | -8.741 | 2.073 | |
| | 1 − 1 | E | 8074.6962(9) | | -8.816 | 0.748 | |
| | 1 − 1 | A | 8074.7526(9) | | -8.816 | 0.748 | |
| $7_{2,5} - 7_{1,6}$ | 7 − 7 | E | 18108.8347(9) | 10.57 | -7.721 | 4.128 | 71.3 |
| | 7 − 7 | A | 18108.9452(9) | | -7.721 | 4.128 | |
| | 8 − 8 | E | 18109.4425(8) | | -7.712 | 4.776 | |
| | 6 − 6 | E | 18109.5301(9) | | -7.714 | 3.634 | |
| | 8 − 8 | A | 18109.5530(8) | | -7.712 | 4.776 | |
| | 6 − 6 | A | 18109.6406(9) | | -7.714 | 3.634 | |
| $4_{0,4} - 3_{1,3}$ | 4 − 3 | A | 21436.9829(8) | 3.11 | -7.711 | 1.525 | 53.1 |
| | 4 − 3 | E | 21437.0118(8) | | -7.711 | 1.525 | |
| | 5 − 4 | A | 21437.2529(8) | | -7.684 | 1.985 | |
| | 5 − 4 | E | 21437.2818(8) | | -7.684 | 1.985 | |
| | 3 − 2 | A | 21437.4648(8) | | -7.721 | 1.160 | |
| | 3 − 2 | E | 21437.4937(8) | | -7.721 | 1.160 | |
| $8_{3,5} - 8_{2,6}$ | 7 − 7 | E | 25851.2367(13) | 14.52 | -7.310 | 3.652 | 38.4 |
| | 9 − 9 | E | 25851.2626(13) | | -7.309 | 4.641 | |
| | 7 − 7 | A | 25851.4598(13) | | -7.310 | 3.652 | |
| | 8 − 8 | E | 25851.4688(13) | | -7.316 | 4.088 | |
| | 9 − 9 | A | 25851.4858(13) | | -7.309 | 4.641 | |
| | 8 − 8 | A | 25851.6919(13) | | -7.316 | 4.088 | |
| $5_{0,5} - 4_{1,4}$ | 5 − 4 | A | 28389.5638(9) | 4.60 | -7.262 | 2.262 | 32.7 |
| | 5 − 4 | E | 28389.5833(9) | | -7.262 | 2.262 | |
| | 6 − 5 | A | 28389.7391(9) | | -7.244 | 2.780 | |
| | 6 − 5 | E | 28389.7587(9) | | -7.244 | 2.780 | |
| | 4 − 3 | A | 28389.8654(9) | | -7.266 | 1.832 | |
| | 4 − 3 | E | 28389.8849(9) | | -7.266 | 1.832 | |
| $7_{0,7} - 6_{1,6}$ | 7 − 6 | A | 41393.0373(12) | 8.45 | -6.678 | 3.810 | 17.4 |
| | 7 − 6 | E | 41393.0406(12) | | -6.678 | 3.810 | |
| | 8 − 7 | A | 41393.0506(12) | | -6.670 | 4.405 | |
| | 8 − 7 | E | 41393.0539(12) | | -6.670 | 4.405 | |
| | 6 − 5 | A | 41393.0999(12) | | -6.680 | 3.292 | |
| | 6 − 5 | E | 41393.1032(12) | | -6.680 | 3.292 | |

Note— The laboratory rotational data of $C_2H_5CONH_2$ were taken from Li et al. (2021). Similar with Table 2, the background temperatures were calculated in accordance to the GBT.

Criteria" (Snyder et al. 2005; Xue et al. 2019) for the identification of a transition of a new interstellar molecule, and as a result, this feature currently remains unidentified.

With the MOLSIM software, a single-excitation-temperature model was used assuming all transitions were thermalized and corrected for optical depth, as discussed in Turner (1991), to estimate an upper limit for the abundance of $C_2H_5CONH_2$ from the PRIMOS data. The upper limit for $C_2H_5CONH_2$ at centimeter wavelengths was determined by the rms noise level from the passband at 41400 MHz, where the absorption line with the strongest predicted intensity was located. In this case, only a single absorption component was considered since no features were detected at any velocity. A similar analysis was performed when reporting an upper limit to the column density of $CH_3CSNH_2$ (Remijan et al. 2022). Assuming the physical conditions of the molecular gas absorbing toward Sgr B2(N) to be similar to those found from the $CH_3CONH_2$ fit (Remijan et al. 2022), i.e. $T_{ex}$ is 5.8 K and the source size is fixed to be 20[11], we determined a $3\sigma$ upper limit of $1.8 \times 10^{14}$ cm$^{-2}$ to the total column density.

The column density upper limits for $C_2H_5CONH_2$ are well constrained from both this work and from (Kolesniková et al. 2022). However, the upper limits to the column density ratios of $C_2H_5CONH_2/NH_2CHO$ and $C_2H_5CONH_2/CH_3CONH_2$ are not - especially from the GBT observations. In general, the column densities of more complex species tend to decrease with increasing molecular complexity. As such, the column density of $C_2H_5CONH_2$ should be lower than the column density of $CH_3CONH_2$ and even lower than the column density of $NH_2CHO$ (See Section 4). This trend is found for the column density ratio $C_2H_5CONH_2/NH_2CHO$. The $C_2H_5CONH_2/NH_2CHO$ ratio upper limit is <0.21 for the GBT observations, and <0.02 and <0.01 from the recent ALMA observations for the N2 and N1S sources, respectively (Kolesniková et al. 2022). However, the sensitivity of the GBT observations cannot additionally constrain the abundance ratio of $C_2H_5CONH_2/CH_3CONH_2$. The upper limit to the $C_2H_5CONH_2/CH_3CONH_2$ ratio is found to be <2.34 from the GBT observations and <0.31 for the source N2 and <0.07 for source N1S from the ALMA analysis. The upper limits from the ALMA observations are factors of 2-3 *lower* than the claimed detection from Li et al. (2021). While not inconsistent with the ALMA upper limits by Kolesniková et al. (2022), the column density ratio of $C_2H_5CONH_2/CH_3CONH_2$ from the GBT observations cannot be used to argue against the claimed detection. Table 3 summarizes the upper limits to the column density ratios found from this work and the recent analysis by (Kolesniková et al. 2022).

## 4. DISCUSSION

### 4.1. *Investigating the Relative Column Density Ratios between -ethyl and -methyl Containing Species.*

Complementary observations taken with facilities operating at different observing frequencies and sensitive to very different spatial scales are essential in trying to determine the overall excitation, column density and distribution of molecular species under differing astronomical conditions. For molecular species similar in structure to $CH_3CONH_2$, namely $CH_3CHO$, $NH_2CHO$ and $(CH_3)_2CO$, all have been detected in both the warm/hot compact molecular core regions toward Sgr B2(N) from ALMA observations (Belloche et al. 2019; Ordu et al. 2019) as well as in the extended, cooler molecular envelopes surrounding the Sgr B2(N) region from GBT observations (Remijan et al. 2022; Hollis et al. 2006). And while there exists orders of magnitude differences in the overall measured total column densities of the species between these two facilities, the relative column density ratios between species are quite similar across observational datasets.

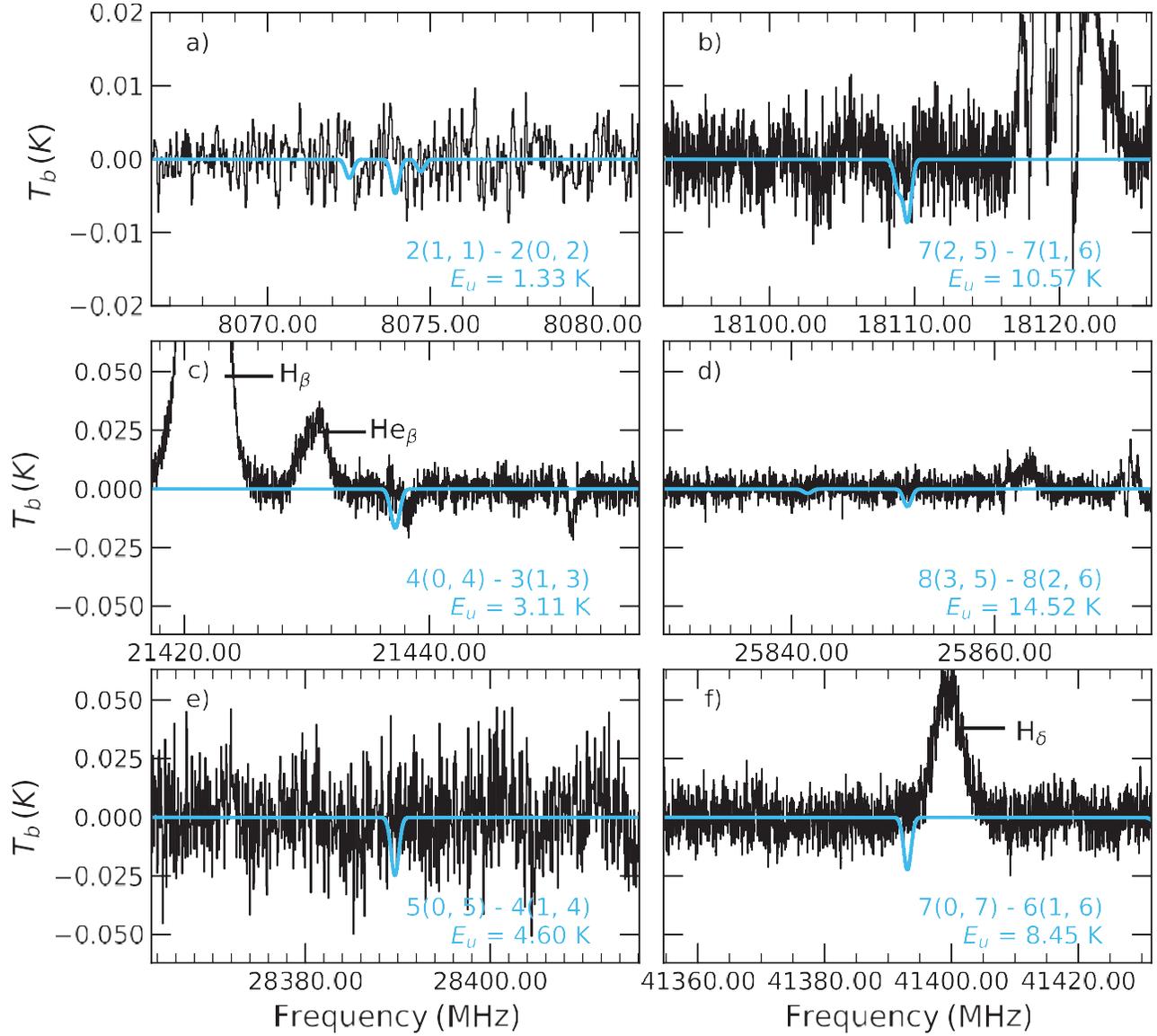

**Figure 2.** The observed and synthetic spectra of $C_2H_5CONH_2$. Simulated $C_2H_5CONH_2$ absorption features (blue trace) overlaid on the PRIMOS observations (black trace) toward Sgr B2(N). A single velocity component of +64 km/s was used to model the transitions. The quantum numbers and upper state energy of each transition are shown. Nearby transitions belonging to other species are labeled. The synthetic spectra for the Sgr B2(N) component (blue trace) were generated from the physical conditions and observing parameters described in Section 3.2, that is, $T_{ex} = 5.8$ K, source size = $20''$, $N_T \sim 1.8 \times 10^{14}$ cm$^{-2}$.

Li et al. (2021) suggest there is an enhanced abundance of $C_2H_5CONH_2$ compared to $CH_3CONH_2$ toward the region Sgr B2(N1E), yet the origin of this enhancement remains unclear. However, the claimed column density ratio between $C_2H_5CONH_2$ and $CH_3CONH_2$ reported in Li et al. (2021) is in general agreement with the relative column density ratios of other -ethyl and -methyl related molecules detected toward a variety of astronomical sources. And is it even possible to find a "global"

**Table 3.** Reported Measurements and Upper Limits of the Column Densities of $C_2H_5CONH_2$ and $CH_3CONH_2$

| Source | Molecule | $T_{ex}$ (K) | $N_T$ (cm$^{-2}$) | Ratio (-$C_2H_5$/-$CH_3$) |
|---|---|---|---|---|
| Sgr B2(N) | $CH_3CONH_2$[a] | 5.8 | $7.7\times10^{13}$ | <2.34 |
| Sgr B2(N) | $C_2H_5CONH_2$ | 5.8 | $<1.8\times10^{14}$ | |
| Sgr B2(N2) | $CH_3CONH_2$[b] | 180 | $1.4\times10^{17}$ | <0.31 |
| Sgr B2(N2) | $C_2H_5CONH_2$ | 180 | $<4.3\times10^{16}$ | |
| Sgr B2(N1S) | $CH_3CONH_2$[b] | 160 | $4.1\times10^{17}$ | <0.07 |
| Sgr B2(N1S) | $C_2H_5CONH_2$ | 160 | $<2.9\times10^{16}$ | |

Note—[a] This work. [b] Column densities and temperatures reported in Kolesniková et al. (2022).

column density ratio between -ethyl and -methyl containing species that may be tied to the relative column densities of the -ethyl and -methyl functional groups?

Finding a consistent relative column density ratio between molecules containing -ethyl and -methyl groups has proven very difficult. The column density of molecules containing an -ethyl group compared to a -methyl group (e.g. ethanol ($CH_3CH_2OH$) vs. methanol ($CH_3OH$)) tend to exhibit smaller abundances for the -ethyl containing species (Belloche et al. 2016) and there are some extreme cases where the differences can be off by several orders of magnitude (Nummelin et al. 2000). These cases are often unique as they may include multiple component fits where the data indicate "core" and "halo" components and a difference in the column densities between these two components or where the column density is determined from optically thick lines. Often the column densities found for some -methyl containing species are determined from the rotational transitions of vibrationally excited state transitions where optical depth effects are typically negligible.

For example, data from Nummelin et al. (2000) showed that the excitation of the $CH_3OH$ lines could only be explained by a 2 component fit toward the Sgr B2(N) region: a "core" component with a temperature of ∼240 K and a total $CH_3OH$ column density of ∼$5.0\times10^{18}$ cm$^{-2}$, and a "halo" component with a temperature of ∼45 K and a total $CH_3OH$ column density of ∼$1.0\times10^{16}$ cm$^{-2}$. In this case, only the "halo" component of $CH_3OH$ was in general agreement with the determined $CH_3CH_2OH$ column density of $4.2\times10^{15}$ cm$^{-2}$ at a temperature of 73 K. However, from the Exploring Molecular Complexity with ALMA (EMoCA) observations taken toward the same region, Belloche et al. (2016) reported a $CH_3OH$ column density of ∼$4.0\times10^{19}$ cm$^{-2}$ and an excitation temperature of 160 K determined from the $v_8$=1 excited state as well as a $CH_3CH_2OH$ column density of ∼$2.0\times10^{18}$ cm$^{-2}$ and an excitation temperature of 150 K.

Observations toward other sources also show the difficulty in determining the column density ratios of -ethyl and -methyl pairs from other molecules. For example, from the ALMA Protostellar Interferometric Line Survey (PILS), the $CH_3CN$ column density was measured to be ∼$8.0\times10^{16}$ cm$^{-2}$ at an excitation temperature of 120 K toward IRAS 16293-2422 source A and a $CH_3CH_2CN$ column density of ∼$9.0\times10^{15}$ cm$^{-2}$ at an excitation temperature of 140 K. Given the large column density and optical depth of the $CH_3CN$ transitions (as was also seen with $CH_3OH$ in the ALMA EMoCA observations) the total column density and temperature were again extrapolated from the $v_8$=1 ex-cited state transitions (Calcutt et al. 2018). However, extrapolation taking into account optical depth effects can also lead to very large differences in measured column densities. Toward NGC 7129, for example, the extrapolated $CH_3CN$ column density was measured to be $\sim 5.2\times10^{18}$ cm$^{-2}$ at a temperature of 400 K, assuming an optical depth as large as 10 from the detected transitions (Fuente et al. 2014). Using only the $v_8$=1 excited state transitions, a total column density closer to $\sim 9.4\times10^{14}$ cm$^{-2}$ at a temperature of 400 K was determined. For $CH_3CH_2CN$, the measured column density was $\sim 5.2\times10^{14}$ cm$^{-2}$ at a temperature of 200 K. As such, reporting an "actual" or "true" column density ratio for any individual source is very difficult, especially using very abundant species towards high density and temperature regions.

It may be possible to better estimate these ratios if we were to consider molecular pairs that may not suffer from such large optical depth effects (e.g. $C_2H_5OCHO$ and $CH_3OCHO$) as would also be the case for the molecules of interest in this study. In addition, to $C_2H_5OCHO$ and $CH_3OCHO$, other molecular pairs such as $C_2H_5CHO$ and $CH_3CHO$, $C_2H_5SH$ and $CH_3SH$ and $C_2H_5NCO$ and $CH_3NCO$ are considered (See Table 4). For the $C_2H_5OCHO$ and $CH_3OCHO$ pair, the measured $C_2H_5OCHO$ column density was determined to be $\sim 5.4\times10^{16}$ cm$^{-2}$ at a temperature of 100 K toward the Sgr B2(N) region (Belloche et al. 2009). From these same observations, the measured $CH_3OCHO$ column density determined was $\sim 4.5\times10^{17}$ cm$^{-2}$ at a temperature of 80 K. Subsequent observations toward the W51 e2 hot molecular core measured a $C_2H_5OCHO$ column density of $\sim 2.0\times10^{16}$ cm$^{-2}$ at a temperature of 78 K and a $CH_3OCHO$ total column density of $\sim 5.0\times10^{17}$ cm$^{-2}$ at a temperature of 112 K (Rivilla et al. 2017).

For the $C_2H_5CHO$ and $CH_3CHO$ pair, observations towards IRAS 16293-2422 B measured a $C_2H_5OCHO$ column density of $\sim 2.2\times10^{15}$ cm$^{-2}$ and a $CH_3CHO$ total column density of $\sim 7.0\times10^{16}$ cm$^{-2}$ at a temperature of 125 K (Lykke et al. 2017). Additionally, the measured $C_2H_5CHO$ column density was determined to be $\sim 7.3\times10^{13}$ cm$^{-2}$ at a temperature of 12 K toward the G+0.693-0.027 region (Sanz-Novo et al. 2022). From these same observations, the measured $CH_3CHO$ column density determined was $\sim 5.0\times10^{14}$ cm$^{-2}$ at a temperature of 9 K. These observations also support the assertion that interstellar aldehydes detected in the extended envelopes are rotationally cold (Hollis et al. 2004b).

As such, considering the relative column density ratios found between lower abundance -ethyl and -methyl containing molecules where transitions are not optically thick, the ratio claimed by Li et al. (2021) is well within the reported ratios found between other species. Table 4 lists the column densities of associated -methyl and -ethyl group species toward this selected sample of sources.

From this limited sample, however, it is not possible to fully extrapolate the relative column densities between -ethyl and -methyl containing molecules (and definitely not the column density ratio between the -ethyl and -methyl functional groups in the ISM) toward astronomical sources. Given the physical and chemical environments where these species are found, such a disparate collection of sources suggests that there is no common relative column density ratio that can be determined between -ethyl and -methyl species. Nevertheless, the supposed high column density found for $C_2H_5CONH_2$ claimed by Li et al. (2021) is not anomalous and furthermore, the proposed formation pathway should not depend on the relative abundance between the -ethyl and -methyl functional groups in the ISM.

*4.2. Formation Route Energetics of $C_2H_5CONH_2$*

**Table 4.** Selected Sample of Measured Abundances of -methyl and -ethyl Containing Molecules

| Source | Molecule | $T_{ex}$ (K) | $N_T$ (cm$^{-2}$) | Ratio (-C$_2$H$_5$/-CH$_3$) |
|---|---|---|---|---|
| Sgr B2(N2) | CH$_3$OH[a] | 160 | 4.0×10$^{19}$ | 0.05 |
| Sgr B2(N2) | C$_2$H$_5$OH | 150 | 2.0×10$^{18}$ | |
| IRAS 16293 A | CH$_3$CN[b] | 120 | 8.0×10$^{16}$ | 0.11 |
| IRAS 16293 A | C$_2$H$_5$CN | 140 | 9.0×10$^{15}$ | |
| NGC 7129 | CH$_3$CN[c] | 400 | 9.4×10$^{14}$ | 0.50 |
| NGC 7129 | C$_2$H$_5$CN | 200 | 5.2×10$^{14}$ | |
| Sgr B2(N) | CH$_3$OCHO[d] | 80 | 4.5×10$^{17}$ | 0.13 |
| Sgr B2(N) | C$_2$H$_5$OCHO | 100 | 5.4×10$^{16}$ | |
| W51 e2 | CH$_3$OCHO[e] | 112 | 5.0×10$^{17}$ | 0.04 |
| W51 e2 | C$_2$H$_5$OCHO | 78 | 2.0×10$^{16}$ | |
| G+0.693 | CH$_3$CHO[f] | 9.4 | 5.0×10$^{14}$ | 0.15 |
| G+0.693 | C$_2$H$_5$CHO | 12 | 7.4×10$^{13}$ | |
| IRAS 16293 A | CH$_3$CHO[g] | 125 | 7.0×10$^{16}$ | 0.03 |
| IRAS 16293 A | C$_2$H$_5$CHO | 125 | 2.2×10$^{15}$ | |
| G+0.693 | CH$_3$SH[h] | ~10 | 6.5×10$^{14}$ | 0.06 |
| G+0.693 | C$_2$H$_5$SH | 10 | 4.2×10$^{13}$ | |
| G+0.693 | CH$_3$NCO[i] | 7.9 | 6.6×10$^{13}$ | 0.12 |
| G+0.693 | C$_2$H$_5$NCO[j] | 10 | 8.1×10$^{12}$ | |

Note—[a] Determined from the $v_8$=1 excited state (Belloche et al. 2016). [b] Determined from the $v_8$=1 excited state (Calcutt et al. 2018). [c] Determined from the $v_8$=1 excited state (Fuente et al. 2014). [d] Column densities and temperatures reported in Belloche et al. (2009). [e] Column densities and temperatures reported in (Rivilla et al. 2017). [f] Column densities and temperatures reported in Sanz-Novo et al. (2022). [g] Column densities and temperatures reported in Lykke et al. (2017). [h] Column densities and temperatures reported in Rodríguez-Almeida et al. (2021a). [i] Column densities and temperatures reported in Zeng et al. (2018). [j] Column densities and temperatures reported in Rodríguez-Almeida et al. (2021b).

Some of the proposed formation routes of C$_2$H$_5$CONH$_2$ as described by Li et al. (2021) suggest pathways driven by -ethyl group addition to stable molecules or radicals. For example, the following reaction was proposed, which is an extension of the Hollis et al. (2006) proposed formation of CH$_3$CONH$_2$ by -methyl group addition:

$$C_2H_5 + NH_2CHO \longrightarrow C_2H_5CONH_2 + H \quad (1)$$

or similarly, reactions of -methyl groups with larger molecules such as CH$_3$CONH$_2$:

$$\mathrm{CH_3 + CH_3CONH_2 \dashrightarrow C_2H_5CONH_2 + H.} \qquad (2)$$

The possible formation routes to $\mathrm{C_2H_5CONH_2}$ proposed by Li et al. (2021) did not contain any information on energetics. As such, to determine their feasibility under interstellar conditions, we investigated the energetics involved in the formation pathways given in Reactions 1 and 2 using the GAUSSIAN 16 suite of programs (Frisch et al. 2016). All geometry optimizations for these calculations were performed using DFT at the B3LYP/6-31G+(d) level of theory (Becke 1993). All transition states found were confirmed using an internal reaction coordinate analysis where the stationary point energies were further constrained at the CCSD(T)/aug-cc-pVTZ level of theory using the optimized geometries and zero point energy corrections from the DFT calculations. Both of the proposed Reactions 1 and 2 were found to be endothermic by 29.52 and 52.19 kJ/mole, respectively. Reaction 1 consists of a two-step formation pathway with energies of activation of 65.36 kJ/mole for the first step and 41.75 kJ/mole for the second. Reaction 2 is a single-step process with a significantly higher activation energy of 259.43 kJ/mole. The potential energy surface for these pathways is shown by Figure 3. Furthermore, additional energetics should be considered for alternate production pathways for the given set of reactants in Reactions 1 and 2. For example, it is possible for the hydrocarbon radicals to abstract a hydrogen from a reacting amide to form the saturated hydrocarbon and an amide radical. Such a reaction pathway was seen to dominate in the gas phase reaction between a -methyl group and $CH_3CONH_2$ or $NH_2CHO$ (Gray & Leyshon 1969) and therefore could also be true for the corresponding reaction between an -ethyl group and $NH_2CHO$. Given this investigation, the proposed formation routes discussed in Li et al. (2021) do not seem to be an efficient way of forming $C_2H_5CONH_2$ given they are largely endothermic and can also form other, more stable products. Further investigations of these reactions on grain surfaces are also necessary to determine their overall efficiency.

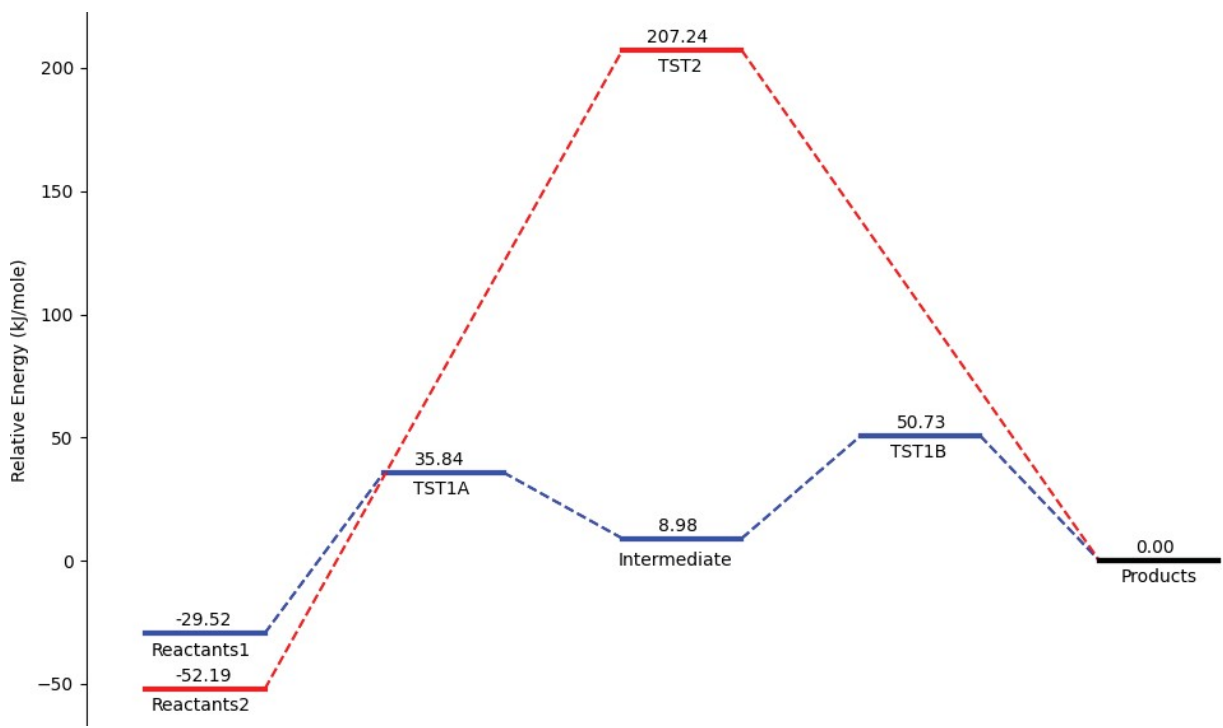

**Figure 3.** Diagram of the potential energy surface for the formation routes to propionamide starting from formamide and the ethyl radical (blue) and acetamide and the methyl radical (red). All stationary points are labelled by their energy in kJ/mole relative to the products, propionamide and a hydrogen atom.

*4.3. Possible New Astronomical Species*

A central unanswered question in astrochemical research is the reason why some molecules containing common functional groups (e.g. -methyl, -amine, -aldehyde, -ethyl groups) are readily found in some astronomical environments and in particular molecular species while others are only detected in specific regions (or not at all). For example, the -ethyl ($-C_2H_5$) functional group is highly abundant and found readily in interstellar cyanides (e.g. $C_2H_5CN$) and alcohols (e.g. $C_2H_5OH$) in various astronomical environments (Coletta et al. 2020), yet relatively limited in interstellar aldehydes (e.g. $C_2H_5CHO$) (Hollis et al. 2004b; Lykke et al. 2017) and esters (e.g. $C_2H_5OCHO$) (Belloche et al. 2009).

It is also interesting from a molecular detection point of view as to why some molecules containing these common functional groups are detected by facilities as disparate as the GBT and ALMA (i.e. present in both the hot compact molecular cores and in the cooler extended outer envelopes). To illustrate this point, Figure 4 shows the "family tree" of detected interstellar aldehydes. Starting with $H_2CO$, one of the earliest polyatomic molecules detected at centimeter wavelengths (Snyder et al. 1969), all species shown in Figure 4 with a green star have been detected toward the Sgr B2(N) region. However, while urea (($NH_2)_2CO$) has been detected with ALMA towards Sgr B2(N), no features of $(NH_2)_2CO$ have been detected in the GBT PRIMOS data. These searches have included both possible absorption line features and transitions that may indicate maser activity. Also, while $(NH_2)_2CO$ was identified by Belloche et al. (2019), it was found toward the source Sgr B2(N1) at a position designated N1S which is slightly offset from the continuum emission peak. No detection of $(NH_2)_2CO$ was made toward the nominal Sgr B2(N2) region where the observations of $(CH_3)_2CO$, $CH_3CONH_2$ and $NH_2CHO$ were measured (Belloche et al. 2017). This possibly suggests unique formation or excitation mechanisms are needed to form and detect $(NH_2)_2CO$ in astronomical environments that is different from the other structurally similar molecular species (e.g. $(NH_2)_2CO$ has also been detected towards G+0.693-0.027 which is a region with quite different physical conditions than Sgr B2(N1)).

Additionally, the column densities of molecules detected with the GBT are orders of magnitude lower than the column densities found of the same molecule detected with ALMA. Using the reported column densities of $CH_3CONH_2$ and $NH_2CHO$ found by ALMA observations (Belloche et al. 2017) as proxies, we find GBT column densities between $10^{3-4}\times$ lower than what was measured by ALMA. As such, given the reported $(NH_2)_2CO$ column density of $N_T = 2.7 \times 10^{16}$ cm$^{-2}$ toward Sgr B2(N1S) (Belloche et al. 2019), if this same ratio holds between the relative column densities toward the extended regions around Sgr B2(N), the possible abundance of $(NH_2)_2CO$ would be $\sim 2.7\times10^{12}$ cm$^{-2}$ or (at best) $\sim 2.7\times10^{13}$ cm$^{-2}$. Similarly, given the claimed column density of $1.5 \times 10^{16}$ cm$^{-2}$ by Li et al. (2021), the possible abundance of $C_2H_5CONH_2$ would be $\sim 1.5\times10^{12}$ cm$^{-2}$ or (at best) $\sim 1.5\times10^{13}$ cm$^{-2}$. This is currently below the detection limit of the PRIMOS survey and, of course, is only to be considered an approximation. In addition, there are numerous reasons as to why molecules are only, and possibly can only be, detected under certain interstellar conditions as the chemical formation routes of interstellar molecules are directly related to the physical conditions of astronomical environments.

Yet, it is interesting to consider the possibility of continuously adding -methyl or -amine groups to form larger molecules that may be present in astronomical environments. Such an illustration is shown in Figure 4, including the 2 proposed formation pathways to $C_2H_5CONH_2$ (boxes) described in Equations 1 and 2. It should be noted that the species shown in Figures 4 and 5 are for illustrative purposes and the energetics involved in the addition of the -methyl or -amine groups need to be carefully investigated (see e.g. Molpeceres & Rivilla (2022) and references therein). While the claimed detection of C$_2$H$_5$CONH$_2$ cannot be verified at this time (Kolesniková et al. 2022, and this work), there are a number of possible molecular species that may be present if their spectra are known.

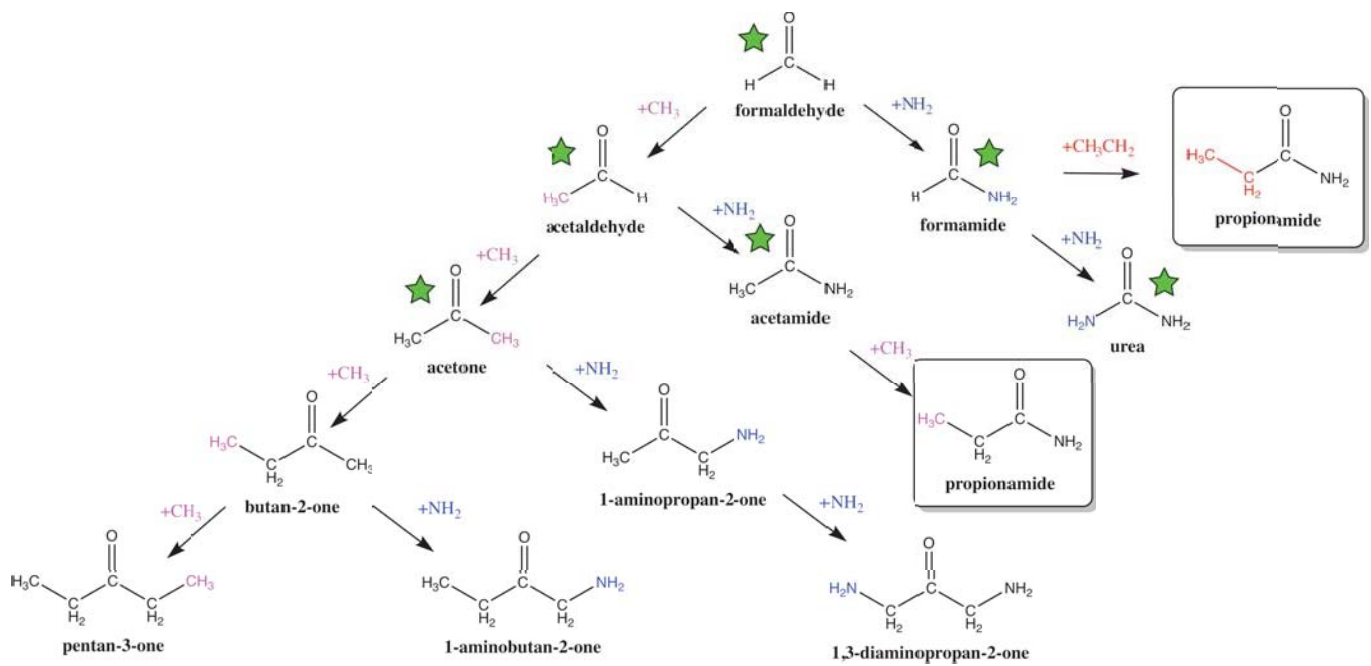

**Figure 4.** Graphical representation of possible molecular structures by subsequent -methyl (CH$_3$) and/or -amine (NH$_2$) addition starting with formaldehyde (H$_2$CO). Note that these are only illustrations of structures and not true representations of possible formation routes.

As previously stated, Figure 4 provides the astronomically detected molecules (green stars) and illustrative pathways by subsequent -methyl and -amine group addition to larger molecules (e.g. C$_2$H$_5$CONH$_2$) that have yet to be confirmed in astronomical sources. For example, if starting with acetone ((CH$_3$)$_2$CO), via subsequent -methyl group additions, it is possible to build up to C$_2$H$_5$COCH$_3$ and (C$_2$H$_5$)$_2$CO). Similarly, subsequent -amine group additions yields CH$_3$COCH$_2$NH$_2$ and (CH$_2$NH$_2$)$_2$CO. And finally, C$_2$H$_5$COCH$_2$NH$_2$ can be achieved via a -methyl and -amine addition from (CH$_3$)$_2$CO.

Of these five molecules, spectroscopic information is available for both C$_2$H$_5$COCH$_3$ (Kroll et al. 2014) and (C$_2$H$_5$)$_2$CO (Nguyen & Stahl 2011). In fact, a limited astronomical search for C$_2$H$_5$COCH$_3$ was conducted toward Orion KL using archival data from the Caltech Submillimeter Observatory (CSO), yet no transitions were detected (Kroll et al. 2012). To date, there have been no astronomical search for (C$_2$H$_5$)$_2$CO and no spectroscopic data currently exists for CH$_3$COCH$_2$NH$_2$, (CH$_2$NH$_2$)$_2$CO, or C$_2$H$_5$COCH$_2$NH$_2$; however, some theoretical work has been done on the molecular structure of (CH$_2$NH$_2$)$_2$CO (von Szentpály et al. 1997).

An extension of this methodology can be applied starting from water (H$_2$O). As shown in Figure 5, by subsequent -methyl and -amine group additions, several common interstellar molecules are identified - specifically methanol (CH$_3$OH) and dimethyl ether ((CH$_3$)$_2$O). Yet, -methyl group addition reactions are often not the preferred formation pathways for these species, such as CH$_3$OH, where the route to form CH$_3$OH is either via subsequent H-addition reactions to CO on the surfaces of interstellar dust grains (Chuang et al. 2016) or more recently, through reactions of CH$_3$O + H$_2$CO (Santos et al. 2022). Subsequent H-addition reactions were also proposed for the formation of both $CH_2CHCHO$ and $CH_3CH_2CHO$ from HCCCHO (Hollis et al. 2004b). As such, the formation of $CH_3CH_2CHO$, for example, does not need to proceed from -ethyl or -methyl addition reactions. It is possible that some of the larger species proposed for astronomical searches could form from more simple molecular precursors via H-addition reactions.

Regardless, unlike the number of species detected in astronomical environments starting from the common precursor $H_2CO$ as shown in Figure 4, fewer species have been identified from Figure 5 (or only recently detected in unique environments such as the detection of $NH_2OH$ toward G+0.693-0.027 (Rivilla et al. 2020)). Molecules such as $CH_3ONH_2$ (Kolesniková et al. 2018) and $(C_2H_5)_2O$ (Medvedev et al. 2004) have measured rotational spectra, but have yet to be detected in astronomical environments[2]. The non-detection of $CH_3ONH_2$ is of particular interest given its relative simplicity, its well characterized laboratory spectrum and its structural similarity to $CH_3CONH_2$ which is now a well known interstellar species that has been detected in both the hot cores and surrounding cold molecular envelope of Sgr B2(N) (Remijan et al. 2022; Hollis et al. 2006; Belloche et al. 2017). However, given the difficulty in of detecting $NH_2OH$, and the fact that it is not present toward the Sgr B2(N) region (Pulliam et al. 2012), if molecules from Figure 5 are to be detected, they may have to be searched for toward other and possibly more unique astronomical environments (such as G+0.693-0.027). Given the new detection of molecules towards G+0.693-0.027 (Rodríguez-Almeida et al. 2021b; Zeng et al. 2021; Rivilla et al. 2022), it appears to exhibit chemistry unique from other hot cores like Sgr B2(N) or Orion KL. In addition, many of the species shown in Figure 5 do not have observed laboratory spectra. As such, Figures 4 and 5 and Table 5 provide possible candidate species for future laboratory and astronomical investigation and summarizes if existing laboratory and astronomical searches have been conducted.

## 5. CONCLUSIONS

Using the PRIMOS data, we report an additional search for $C_2H_5CONH_2$ at low frequencies carried out with the GBT. As a result of this search, no spectral features of $C_2H_5CONH_2$ were detected. From this set of observations, an upper limit for $C_2H_5CONH_2$ at centimeter wavelengths was determined to be $N_T < 1.8 \times 10^{14}$ cm$^{-2}$. This upper limit was set by using the same physical parameters of Sgr B2(N) found from the $CH_3CONH_2$, where $T_{ex} = 5.8$ K and a molecular source size = 20[11]. Taking into account the frequency-dependent source sizes of both continuum emission and molecular gas, the model with a single excitation temperature of 5.8 K and a total column density of $8.5 \times 10^{14}$ cm$^{-2}$ matches the observed absorption features of $NH_2CHO$ toward Sgr B2(N). The determined column density was very close to the determination made by Hollis et al. (2006) based on the range of temperatures used in that work and until sufficient modeling is performed to account for the emission detected by maser activity of the other detected $NH_2CHO$ lines, this will remain the best column density determined at a $T_{ex} = 5.8$ K to characterize the spectrum of $NH_2CHO$ at frequencies observed with the GBT.

This work again calls into question the initial detection of $C_2H_5CONH_2$ by Li et al. (2021) and indicates that more complex peptide-like structures may have difficulty forming in the ISM and/or are below the detection limits of current astronomical facilities. In this work, several possible new species were identified as potential astronomical molecules given their structural similarity with al-

---

[2] It should be noted that Kuan et al. (1999) claimed a tentative detection of both $(C_2H_5)_2O$ and $CH_3OC_2H_5$ using the NRAO 12-m that, as of yet, have not been verified.

**Table 5.** Summary of the Existing Laboratory Data or Astronomical Searches Conducted of Potential Astronomical Molecules

| Molecular Formula | Spectrum Known? | Astronomical Search? | References |
| --- | --- | --- | --- |
| $C_2H_5CONH_2$ | **YES** | **YES** | Li et al. (2021), this work |
| $CH_3COCH_2NH_2$ | NO | NO | |
| $(CH_2NH_2)_2CO$ | NO | NO | |
| $C_2H_5COCH_3$ | **YES** | **YES** | Kroll et al. (2014, 2012) |
| $(C_2H_5)_2CO$ | **YES** | NO | Nguyen & Stahl (2011) |
| $NH_2CH_2COCH_2CH_3$ | NO | NO | |
| $(NH_2)_2O$ | NO | NO | |
| $CH_3ONH_2$ | **YES** | **YES** | Kolesniková et al. (2018) |
| $CH_3OCH_2NH_2$[a] | NO | NO | |
| $(NH_2CH_2)_2O$ | NO | NO | |
| $C_2H_5OCH_3$[b] | **YES** | **YES** | Tercero et al. (2015) |
| $(C_2H_5)_2O$ | **YES** | NO | Medvedev et al. (2004) |
| $C_2H_5OCH_2NH_2$ | NO | NO | |

Note—References for laboratory and astronomical searches located in the manuscript text. [a] Methoxymethanimine ($CH_3OCH_2NH_2$) is a structural isomer of aminoethanol ($NH_2CH_2CH_2OH$) - a molecule with well characterized laboratory spectrum (Widicus et al. 2003). [b] A tentative detection of $C_2H_5OCH_3$ has been reported in Tercero et al. (2015)

ready known molecules. For several of these molecules, e.g. $CH_3ONH_2$, $(C_2H_5)_2O$, and $C_2H_5COCH_3$, their rotational spectra have already been characterized. For other molecules such as $CH_3COCH_2NH_2$ and $(CH_2NH_2)_2CO$, laboratory data is still needed before an astronomical search can be conducted. While structural similarity with previously detected species may be an adequate proxy for a dedicated astronomical search, especially for those species that contain abundant functional groups such as -ethyl, methyl and -amine groups, understanding the chemical pathways for large molecule production in astronomical environments should be the goal in motivating future astronomical molecular observations.

We would like to thank the anonymous referee who favorably reviewed and provided valuable feedback that greatly improved the manuscript. This paper makes use of the PRIMOS data under GBT Archive Project Code AGBT07A-051. The Green Bank Observatory is a facility of the National Science Foundation operated under cooperative agreement by Associated Universities, Inc. The National Radio Astronomy Observatory is a facility of the National Science Foundation operated under cooperative agreement by Associated Universities, Inc. C. Xue was supported for portions of this work by a grant from the MIT Research Support Committee and by an NRAO Grote Reber Fellowship.

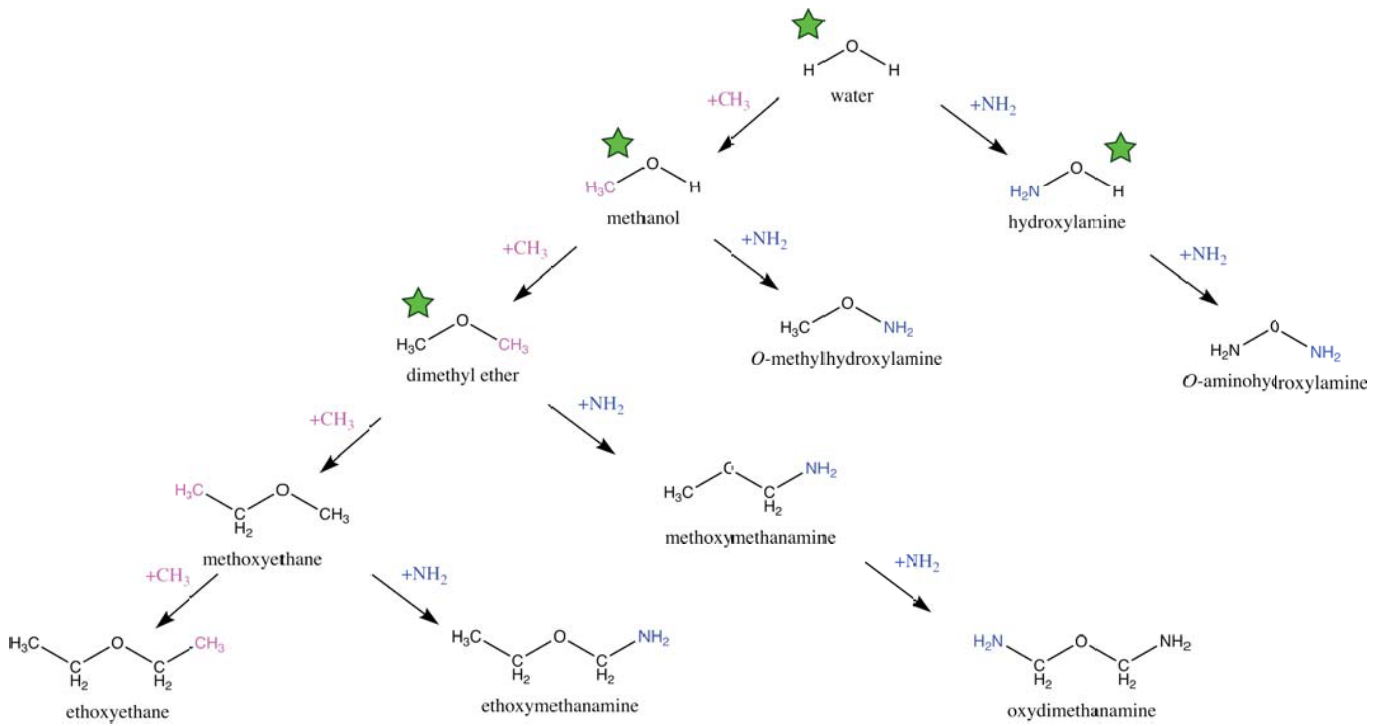

**Figure 5.** Graphical representation of possible molecular structures by subsequent -methyl ($CH_3$) and/or -amine ($NH_2$) addition starting with water ($H_2O$). Caveat concerning possible formation routes is the same as in Figure 4.

## REFERENCES

Adande, G. R., Woolf, N. J., & Ziurys, L. M. 2013, Astrobiology, 13, 439, doi: 10.1089/ast.2012.0912

Becke, A. D. 1993, Journal of Chemical Physics, 98, 5648, doi: 10.1063/1.464913

Belloche, A., Garrod, R. T., Müller, H. S. P., et al. 2009, A&A, 499, 215, doi: 10.1051/0004-6361/200811550

—. 2019, A&A, 628, A10, doi: 10.1051/0004-6361/201935428

Belloche, A., Müller, H. S. P., Garrod, R. T., & Menten, K. M. 2016, A&A, 587, A91, doi: 10.1051/0004-6361/201527268

Belloche, A., Meshcheryakov, A. A., Garrod, R. T., et al. 2017, A&A, 601, A49, doi: 10.1051/0004-6361/201629724

Bisschop, S. E., Jørgensen, J. K., van Dishoeck, E. F., & de Wachter, E. B. M. 2007, A&A, 465, 913, doi: 10.1051/0004-6361:20065963

Blake, G. A., Sutton, E. C., Masson, C. R., & Phillips, T. G. 1986, ApJS, 60, 357, doi: 10.1086/191090

Calcutt, H., Jørgensen, J. K., Müller, H. S. P., et al. 2018, A&A, 616, A90, doi: 10.1051/0004-6361/201732289

Candelaria, T. M., Meier, D. S., Ott, J., & Mills, E. A. C. 2021, in Astronomical Society of the Pacific Conference Series, Vol. 528, New Horizons in Galactic Center Astronomy and Beyond, ed. M. Tsuboi & T. Oka, 113

Chuang, K. J., Fedoseev, G., Ioppolo, S., van Dishoeck, E. F., & Linnartz, H. 2016, MNRAS, 455, 1702, doi: 10.1093/mnras/stv2288

Coletta, A., Fontani, F., Rivilla, V. M., et al. 2020, A&A, 641, A54, doi: 10.1051/0004-6361/202038212

Colzi, L., Rivilla, V. M., Beltrán, M. T., et al. 2021, A&A, 653, A129, doi: 10.1051/0004-6361/202141573

Corby, J. F., Jones, P. A., Cunningham, M. R., et al. 2015, MNRAS, 452, 3969, doi: 10.1093/mnras/stv1494

Faure, A., Remijan, A. J., Szalewicz, K., & Wiesenfeld, L. 2014, ApJ, 783, 72, doi: 10.1088/0004-637X/783/2/72

Frisch, M. J., Trucks, G. W., Schlegel, H. B., et al. 2016, Gaussian~16 Revision C.01

Fuente, A., Cernicharo, J., Caselli, P., et al. 2014, A&A, 568, A65,
doi: 10.1051/0004-6361/201323074

Gray, P., & Leyshon, L. J. 1969, Transactions of the Faraday Society, 65, 780,
doi: 10.1039/TF9696500780

Halfen, D. T., Ilyushin, V., & Ziurys, L. M. 2011, ApJ, 743, 60, doi: 10.1088/0004-637X/743/1/60

Hollis, J. M., Jewell, P. R., Lovas, F. J., & Remijan, A. 2004a, ApJL, 613, L45,
doi: 10.1086/424927

Hollis, J. M., Jewell, P. R., Lovas, F. J., Remijan, A., & Møllendal, H. 2004b, ApJL, 610, L21,
doi: 10.1086/423200

Hollis, J. M., Jewell, P. R., Remijan, A. J., & Lovas, F. J. 2007, ApJL, 660, L125,
doi: 10.1086/518124

Hollis, J. M., Lovas, F. J., Remijan, A. J., et al. 2006, ApJL, 643, L25, doi: 10.1086/505110

Jiménez-Serra, I., Martín-Pintado, J., Rivilla, V., et al. 2020, Astrobiology, 20,
doi: 10.1089/ast.2019.2125

Kolesniková, L., Tercero, B., Alonso, E. R., et al. 2018, A&A, 609, A24,
doi: 10.1051/0004-6361/201730744

Kolesniková, L., Belloche, A., Koucký, J., et al. 2022, A&A, 659, A111,
doi: 10.1051/0004-6361/202142448

Kroll, J. A., Shipman, S. T., & Widicus Weaver, S. L. 2014, Journal of Molecular Spectroscopy, 295, 52, doi: 10.1016/j.jms.2013.10.005

Kroll, J. A., Weaver, S. L. W., & Shipman, S. T. 2012, in 67th International Symposium on Molecular Spectroscopy, RF06

Kuan, Y. J., Charnley, S. B., Wilson, T. L., et al. 1999, 194, 71.02

Lee, K. L. K., Loomis, R. A., Xue, C., El-Abd, S., & McGuire, B. A. 2021, molsim, 0.3.0, Zenodo,
doi: 10.5281/zenodo.5497790

Li, J., Wang, J., Lu, X., et al. 2021, ApJ, 919, 4,
doi: 10.3847/1538-4357/ac091c

Ligterink, N. F. W., El-Abd, S. J., Brogan, C. L., et al. 2020, ApJ, 901, 37,
doi: 10.3847/1538-4357/abad38

Liu, S.-Y., Mehringer, D. M., & Snyder, L. E. 2001, ApJ, 552, 654, doi: 10.1086/320563

Loomis, R. A., Zaleski, D. P., Steber, A. L., et al. 2013, ApJL, 765, L9,
doi: 10.1088/2041-8205/765/1/L9

López-Sepulcre, A., Balucani, N., Ceccarelli, C., et al. 2019, ACS Earth and Space Chemistry, 3, 2122, doi: 10.1021/acsearthspacechem.9b00154

Lykke, J. M., Coutens, A., Jørgensen, J. K., et al. 2017, A&A, 597, A53,
doi: 10.1051/0004-6361/201629180

Marstokk, K. M., Møllendal, H., & Samdal, S. 1996, Journal of Molecular Structure, 376, 11,
doi: 10.1016/0022-2860(95)09045-2

McGuire, B. A. 2022, ApJS, 259, 30,
doi: 10.3847/1538-4365/ac2a48

McGuire, B. A., Loomis, R. A., Charness, C. M., et al. 2012, ApJL, 758, L33,
doi: 10.1088/2041-8205/758/2/L33

Medvedev, I., Winnewisser, M., De Lucia, F. C., et al. 2004, Journal of Molecular Spectroscopy, 228, 314, doi: 10.1016/j.jms.2004.06.011

Mehringer, D. M., Snyder, L. E., Miao, Y., & Lovas, F. J. 1997, ApJL, 480, L71,
doi: 10.1086/310612

Molpeceres, G., & Rivilla, V. M. 2022, arXiv e-prints, arXiv:2206.00350.
https://arxiv.org/abs/2206.00350

Müller, H. S. P., Schlöder, F., Stutzki, J., & Winnewisser, G. 2005, Journal of Molecular Structure, 742, 215,
doi: 10.1016/j.molstruc.2005.01.027

Neill, J. L., Muckle, M. T., Zaleski, D. P., et al. 2012, ApJ, 755, 153,
doi: 10.1088/0004-637X/755/2/153

Nguyen, H. V. L., & Stahl, W. 2011, ChemPhysChem, 12, 1900,
doi: https://doi.org/10.1002/cphc.201001021

Nummelin, A., Bergman, P., Hjalmarson, Å., et al. 2000, ApJS, 128, 213, doi: 10.1086/313376

Ordu, M. H., Zingsheim, O., Belloche, A., et al. 2019, A&A, 629, A72,
doi: 10.1051/0004-6361/201935887

Pulliam, R. L., McGuire, B. A., & Remijan, A. J. 2012, ApJ, 751, 1,
doi: 10.1088/0004-637X/751/1/1

Remijan, A., Xue, C., Margulès, L., et al. 2022, A&A, 658, A85,
doi: 10.1051/0004-6361/202142504

Rivilla, V. M., Beltrán, M. T., Martín-Pintado, J., et al. 2017, A&A, 599, A26,
doi: 10.1051/0004-6361/201628823

Rivilla, V. M., Martín-Pintado, J., Jiménez-Serra, I., et al. 2020, ApJL, 899, L28, doi: 10.3847/2041-8213/abac55

Rivilla, V. M., Colzi, L., Jiménez-Serra, I., et al. 2022, ApJL, 929, L11, doi: 10.3847/2041-8213/ac6186

Rodríguez-Almeida, L. F., Jiménez-Serra, I., Rivilla, V. M., et al. 2021a, ApJL, 912, L11, doi: 10.3847/2041-8213/abf7cb

Rodríguez-Almeida, L. F., Rivilla, V. M., Jiménez-Serra, I., et al. 2021b, A&A, 654, L1, doi: 10.1051/0004-6361/202141989

Rubin, R. H., Swenson, G. W., J., Benson, R. C., Tigelaar, H. L., & Flygare, W. H. 1971, ApJL, 169, L39, doi: 10.1086/180810

Santos, J. C., Chuang, K.-J., Lamberts, T., et al. 2022, ApJL, 931, L33, doi: 10.3847/2041-8213/ac7158

Sanz-Novo, M., Belloche, A., Rivilla, V. M., et al. 2022, arXiv e-prints, arXiv:2203.07334. https://arxiv.org/abs/2203.07334

Snyder, L. E., Buhl, D., Zuckerman, B., & Palmer, P. 1969, PhRvL, 22, 679, doi: 10.1103/PhysRevLett.22.679

Snyder, L. E., Lovas, F. J., Hollis, J. M., et al. 2005, ApJ, 619, 914, doi: 10.1086/426677

Tercero, B., Cernicharo, J., López, A., et al. 2015, A&A, 582, L1, doi: 10.1051/0004-6361/201526255

Thiel, V., Belloche, A., Menten, K. M., Garrod, R. T., & Müller, H. S. P. 2017, A&A, 605, L6, doi: 10.1051/0004-6361/201731495

Turner, B. E. 1991, ApJS, 76, 617, doi: 10.1086/191577

von Szentpály, L., Shamovsky, I. L., Ghosh, R., & Dakkouri, M. 1997, The Journal of Physical Chemistry A, 101, 3032, doi: 10.1021/jp963079w

Widicus, S. L., Drouin, B. J., Dyl, K. A., & Blake, G. A. 2003, Journal of Molecular Spectroscopy, 217, 278, doi: https://doi.org/10.1016/S0022-2852(02)00056-5

Winnewisser, G., & Churchwell, E. 1975, ApJL, 200, L33, doi: 10.1086/181890

Xue, C., Remijan, A. J., Burkhardt, A. M., & Herbst, E. 2019, ApJ, 871, 112, doi: 10.3847/1538-4357/aaf738

Zeng, S., Jiménez-Serra, I., Rivilla, V. M., et al. 2018, MNRAS, 478, 2962, doi: 10.1093/mnras/sty1174

—. 2021, ApJL, 920, L27, doi: 10.3847/2041-8213/ac2c7e